\documentclass[12pt,twoside]{article}     

\usepackage{graphicx}   
\usepackage{amsmath}     
\usepackage{amssymb}     
\usepackage{amsfonts}    
\usepackage{amsthm}      

\newtheoremstyle{theorem}
  {14pt}
  {14pt}
  {}
  {}
  {\bfseries}
  {}
  {0.5em}
  {}
  
\newtheoremstyle{defnsty}
   {14pt}
  {14pt}
  {}
  {}
  {\bfseries}
  {}
  {\newline}
  {}

\theoremstyle{theorem}
\newtheorem{thm}{Theorem}[]
\theoremstyle{theorem}
\newtheorem{cor}{Corollary}[]
\theoremstyle{defnsty}
\newtheorem{defn}{Definition}[]

\theoremstyle{theorem}
\newtheorem{ex}{Example}[]

\newcommand{\journal}[7]{#1 (#2). #3, \textit{#4}, \textbf{#5}, #6-#7.}

\newcommand{\proc}[8]{#1 (#2). #3, in: #4, eds. #5, #6, #7.}

\newcommand{\book}[5]{#1 (#2). \textit{#3}, #4, #5.}

\newcommand{\phd}[4]{#1 (#2). \textit{#3}, Ph.D. dissertation, #4.}

\newcommand{\JGP}{J. Geom. Phys.}
\newcommand{\AMS}{American Mathematical Society}

\newcommand{\MPCPS}{Math. Proc. Camb. Phil. Soc.}

\newcommand{\thereals}{\ensuremath{\mathbb{R}}}

\newcommand{\manifold}[1]{\ensuremath{\mathcal{#1}}}
\newcommand{\spacetime}[2]{\ensuremath{(\manifold{#1},#2)}}
\newcommand{\curvefamily}[1]{\ensuremath{\mathcal{#1}}}
\newcommand{\set}[1]{\ensuremath{\mathcal{#1}}}
\newcommand{\bset}[1]{\ensuremath{#1}}
\newcommand{\cmptset}[1]{\ensuremath{#1}}

\newcommand{\covers}{\ensuremath{\triangleright}}

\newcommand{\goesto}{\ensuremath{\longrightarrow}}
 
\renewcommand{\mapsto}{\rightarrow}

\newcommand{\nbhd}[1]{\ensuremath{\mathcal{#1}}} 
\newcommand{\atlas}[1]{\ensuremath{\mathcal{#1}}}
\newcommand{\chartset}[2]{\ensuremath{\mathcal{#1}_{#2}}}
\newcommand{\cover}[2]{\ensuremath{ \{\chartset{#1}{#2} \}}}

\newcommand{\abndry}[1]{\ensuremath{\mathcal{B}(\manifold{#1})}} 
\newcommand{\envbndry}[2]{\ensuremath{\partial_{#1}\manifold{#2}}} 
\newcommand{\envelopment}[3]{\ensuremath{(\manifold{#1},\manifold{#2},#3)}}

\newcommand{\extension}[5]{\ensuremath{(\manifold{#1},#2,\manifold{#3},#4,#5)}}

\begin{document} 

\title{The Stability of Abstract Boundary Essential Singularities}

\author{Michael J. S. L. Ashley\footnote{Department of Physics and Theoretical 
Physics, Faculty of Science, The Australian National University, 
Canberra ACT 0200, Australia. Email: \texttt{mike@einstein.anu.edu.au}, Cell Phone: 
+61 402 356554, Fax: +61 2 6125 0741}}  

\date{18 October 2001}

\maketitle

%
%
%
%

\abstract{The abstract boundary has, in recent years, proved a general 
and flexible way to define the singularities of space-time.
In this approach an essential singularity is a non-regular boundary 
point of an embedding which is accessible by a chosen family of curves 
within finite parameter distance.
Ashley and Scott proved the first theorem relating essential 
singularities in strongly causal space-times to causal geodesic 
incompleteness.
Linking this with the work of Beem on the $C^{r}$-stability of 
geodesic incompleteness allows proof of the stability of these 
singularities.
Here I present this result stating the conditions under which essential 
singularities are $C^{1}$-stable against perturbations of the metric.

%
%
%
%

\textbf{Keywords:} abstract boundary, essential singularity, stability, space-time

%
%
%
%

\section{Introduction} The stability of the physical features of 
space-time has been a significant area of inquiry since the production
of the first exact solutions of the Einstein equation.
It has always been thought that for a given space-time to be physically 
reasonable, or for a given feature to exist in the universe at large, 
that the space-time be robust against perturbations of the metric.
The issue of stability has, however, been somewhat difficult to define 
in the abstract sense of a pseudo-Riemannian manifold since there has never 
been a completely coordinate invariant method of defining metrics that 
are \emph{near} one another.
In the practical mathematical and geometrical sense the difficulty has 
arisen because at present there is no candidate for a topology on the 
space of metrics over a given manifold preserving coordinate invariance, 
which is a crucial feature of general relativity.  
Nevertheless with strict conditions on the allowed coverings of 
coordinate charts for a space-time it is possible to derive some 
important results.
With this in mind I will use the Whitney $C^{r}$-fine topology on the 
space of metrics and restate the notion of $C^{r}$-stability for some 
feature of a space-time.
I will then proceed to present some physically intuitive examples of the 
use of the $C^{r}$-fine topology with a view to summarising the literature 
on the stability of geodesic completeness/incompleteness relevant to 
creating a stability theorem for abstract singularities.
Finally I will quote the result of Ashley and Scott and present the 
stability theorem for abstract boundary singularities.

%
%
%
%

\section{A review of the Whitney $C^{r}$-fine topology on the space of metrics}
Let $\atlas{A}=\cover{U}{i}$ be a chosen fixed covering of 
\manifold{M} by a countable collection of charts of \manifold{M}.
We will also assume that every chart has compact closure in a larger 
chart, (i.e. for all $i$, $\overline{\chartset{U}{i}}$ is compact, and there exists a 
\chartset{V}{j} so that $\overline{\chartset{U}{i}} \subset \chartset{V}{j}$, 
with $\cover{V}{j}$ forming an atlas on \manifold{M}) and that the 
covering of \manifold{M} by the \chartset{U}{i} is locally finite.
Now let $\epsilon:\manifold{M}\mapsto\thereals^{+}$ be a continuous 
function.
\begin{defn}
    For any two Lorentzian metrics $g$, $h$, we write,
    \begin{equation}
        |g-h|_{r,\set{P}}<\epsilon,
    \end{equation}
    if for each point $p \in \set{P} \subset \manifold{M}$,
    \begin{equation}
	|g_{ab}-h_{ab}|<\epsilon(p)\text{ and }|g_{ab,c_{1}c_{2} \ldots 
     c_{r}}-h_{ab,c_{1}c_{2} \ldots c_{r}}|<\epsilon(p)
    \end{equation}
    when the metric is evaluated at $p$ for all indices $a$, $b$, $c_{1}$, 
    $c_{2}$, \ldots, $c_{r}$ in all the given charts $\chartset{U}{i} \in 
    \atlas{A}$ which contain $p$.
\end{defn}
\begin{defn}[Whitney $C^{r}$-fine topologies]
    The \emph{Whitney $C^{r}$-fine topologies} (or simply the 
    \emph{$C^{r}$-fine topologies}) are defined by basis neighbourhoods of the 
    form
    \begin{equation}
	\nbhd{N}(g,\epsilon):=\{h: |g-h|_{r,\manifold{M}}<\epsilon\}
    \end{equation}
    about each metric $g$ and for each continuous function 
    $\epsilon:\manifold{M}\mapsto\thereals^{+}$.
\end{defn}
If $h \in \nbhd{N}(g,\epsilon)$ for some given metric $g$, then the two metrics $g$ and $h$ are termed 
$C^{r}$-\emph{close}.

\begin{defn}[$C^{r}$-stable property]
    A property, $X$, of a space-time, \spacetime{\manifold{M}}{g}, is 
    termed $C^{r}$-stable if it is true for every metric in some open 
    neighbourhood of $g$ in the $C^{r}$-fine topology.
\end{defn}

The $C^{r}$-fine topology may be shown to be independent of the cover 
$\cover{U}{i}$ if the conditions above are satisfied.
However, it is worthwhile noting how these conditions are used to 
restrict the allowed coverings.
For example, if the covering is not locally finite then it is possible 
that there would be no partition of unity available over the charts to 
guarantee consistent definition of the metric. 
It follows that $\epsilon$-neighbourhoods of that metric would be ill-defined.
Similarly if one could not guarantee that a chart had compact closure 
in a chart of another atlas then $\epsilon$ may not possess a maximum.
This result would then make it impossible to guarantee that the metrics 
and/or their derivatives would have their deviation confined.

It is also worthwhile noting that coverings of this sort exist for all 
except the most pathological of examples and so these conditions do not 
really pose much of an impediment to the practical use of the $C^{r}$-fine 
topology.
Consequently, at present, the $C^{r}$-fine topology over metrics is 
arguably the most straightforward and practical notion for the \emph{nearness} 
of metrics.

%
%
%
%

\section{A physical interpretation of the $C^{r}$-fine topologies}

I will now digress to give an intuitive description of the  
$C^{r}$-fine topology.
For the cases $r=0, 1, 2$ it is relatively simple to visualise the physical 
relationship between $C^{r}$-close metrics.
\begin{ex}
    If two metrics $g$, $h$ for the space-times 
    \spacetime{\manifold{M}}{g} and \spacetime{\manifold{M}}{h} are 
    $C^{0}$-close, then their metric components are close, implying that 
    the light cones of equivalent points in both space-times are 
    \emph{close}.
\end{ex}
\begin{ex}
    If two metrics $g$, $h$ for the space-times 
    \spacetime{\manifold{M}}{g} and \spacetime{\manifold{M}}{h} are 
    $C^{1}$-close, then the metric components and their first 
    derivatives (and hence the Christoffel symbol functions on 
    \manifold{M}, ${\Gamma^{a}}_{bc}(x)$) are close.
    This implies, by the continuous depedence of the solutions of the 
    geodesic equation on the Christoffel symbols, that the 
    geodesic systems under both metrics are \emph{close} in addition to the light 
    cone structure.
    One should consult Beem, Ehrlich and Easley \cite[p. 247]{BEE.book}
    for additional references.
\end{ex}
\begin{ex}
    If two metrics $g$, $h$ are $C^{2}$-close, then additionally 
    we have that the second derivatives of the metric are close for 
    both metrics and hence the components of the Riemann curvature 
    tensor and other Riemann derived objects (e.g. curvature 
    invariants, Ricci, $R_{ab}$, and Weyl tensors, $C_{abcd}$) are 
    \emph{close}.
\end{ex}

One should now also be able to extrapolate the above examples to 
higher derivatives.
For example, if two metrics are $C^{3}$-close then we would expect 
that in addition to the light cones, geodesic systems and curvature 
being close that the first derivatives of the Riemann tensor would 
also be close.
By analogy we can produce interpretations for $C^{r}$-fine topologies 
with larger $r$.
Note that as $r$ increases, more metrics are excluded from any given 
$\epsilon$-neighbourhood, $\nbhd{N}(g,\epsilon)$, and the resulting 
topologies about the given metric are \emph{finer}.
Correspondingly any property proved to be $C^{s}$-stable will also be 
$C^{r}$-stable, whenever $s\leqslant r < \infty$.
It is also important to remember that the $C^{r}$-fine topologies are 
in a certain sense too coarse since they also include too many metrics 
which one may not wish to consider close\footnote{The metrics allowed 
in these $C^{r}$-fine neighbourhoods may correspond to non-physical 
curvature sources.
A more detailed discussion follows in \S \ref{stability.sec}.}.
For example, one may choose $\epsilon$ functions which are very small 
in some compact region of \manifold{M} but are far from zero elsewhere.
The resulting $C^{r}$-fine open neighbourhood, $\nbhd{N}(g,\epsilon)$, 
will contain not only metrics whose values and derivatives are close 
to those of $g$ everywhere in \manifold{M} but also those that deviate 
wildly outside of the compact region.
An example of this behaviour is presented in Figure \ref{C0fine.fig}.
\begin{figure}[ht!]
    \centering
    \includegraphics{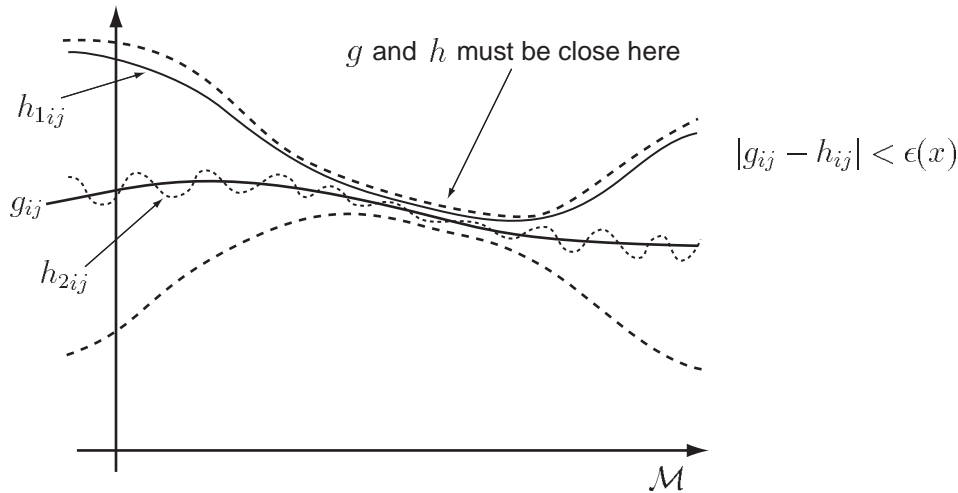}
    \caption[An example of the deviations allowed in the 
    $C^{0}$-fine topology on metrics.]{The figure compares the values 
    of a metric component from $g$ and two $C^{0}$-close metrics, 
    $h_{1}$ and $h_{2}$.
    Note that the metric component, ${h_{1}}_{ij}$, can vary 
    significantly from $g_{ij}$ since $\epsilon$ is only small in a compact 
    region.
    The metric component, ${h_{2}}_{ij}$, has values $\epsilon$-close to 
    $g_{ij}$ but has a wildly varying first derivative.
    Hence the topology allows both ${h_{1}}_{ij}$ and ${h_{2}}_{ij}$ to 
    be close to $g_{ij}$ even though it appears that only ${h_{2}}_{ij}$ 
    should be included.
    Consequently more metrics are present in $\nbhd{N}(g,\epsilon)$ than one's 
    intuition might indicate.
    One may also want for ${h_{2}}_{ij}$ to be excluded from this 
    $\epsilon$-neighbourhood due to its wildly deviating derivative.
    This metric could be eliminated by choosing the topology to be 
    $C^{1}$-fine, since then the neighbourhood would not contain metrics where 
    the slope differed more than $\epsilon$ from that of $g_{ij}$.
    One can, however, easily devise examples where the second 
    and higher derivatives behave pathologically.
    Seeking greater values of $r$ for the $C^{r}$-fine topologies 
    would lead to even finer topologies and exclude these cases.}
    \label{C0fine.fig}
\end{figure}

An important application of the $C^{r}$-fine topology can be seen in the analysis 
of the causal structure of space-times.
Using the above described notion that if two Lorentzian metrics 
are $C^{0}$-close, then their light cones are close, we 
obtain the following precise definition of the stable causality 
property for space-time (as motivated by Geroch 
\cite[p. 241]{Gerochglobalanalysis.paper}):

\begin{defn}[stable causality]
    A space-time \spacetime{\manifold{M}}{g} is \emph{stably causal}  
    if there exists a $C^{0}$-fine neighbourhood, $U(g)$, of the 
    Lorentzian metric, $g$, such that for each $h \in U(g)$, 
    \spacetime{\manifold{M}}{h} is causal.
    \label{stablecausality.defn}
\end{defn}

Hence stably causal space-times remain causal under small 
$C^{0}$-fine perturbations of the metric.
The reader should note that Definition \ref{stablecausality.defn} (from 
Beem, Ehrlich and Easley \cite[p. 63]{BEE.book}) redefines Geroch's idea 
of `the spreading of light cones' precisely and the interested reader is 
asked to compare this with the alternate definition given in Hawking and 
Ellis \cite[p. 198]{HE.book}.

\subsection{The stability of geodesic completeness/incompleteness}

The stability of geodesic completeness/incompleteness has, over the 
years, been investigated closely by Beem and Ehrlich \cite{BeemEhrlich.paper} 
and also Williams \cite{Williams.thesis}.
A thorough review of the literature on the stability of completeness 
and incompleteness is provided in Beem, Ehrlich and Easley \cite[p. 
239-270]{BEE.book}.

Examples by Williams \cite{Williams.thesis} show that both geodesic 
completeness and geodesic incompleteness are not $C^{r}$-stable for 
space-time, 
in general.
In addition, Williams also provided examples showing that these properties 
may fail to be stable even for compact/non-compact space-times.
Of course this still leaves the possibility that with additional 
constraints made on the space-time that geodesic completeness/incompleteness 
may be stable.
Beem showed that geodesic incompleteness is, in fact, $C^{1}$-stable 
for strongly causal space-times.
This work is relevant to our task and the following presentation is 
designed to guide the reader unfamiliar with this result.

I will need precise notions of imprisonment and partial imprisonment 
for curves in space-time since partial imprisonment is very closely 
related to the strong causality condition on a space-time.
The following definitions are those used by Beem in the proof of the 
stability theorems.
\begin{defn}[Imprisonment and Partial Imprisonment]
    Let $\gamma:(a,b)\mapsto\manifold{M}$ be an inextendible geodesic.
    \begin{enumerate}
	\item The geodesic, $\gamma$, is \emph{partially imprisoned} as $t 
	\goesto b$ if there is a compact set $\cmptset{K}\subseteq\manifold{M}$ 
	and a sequence $\{x_{i}\}$ with $x_{i} \goesto b$ from below such 
	that $\gamma(x_{i}) \in \cmptset{K}$ for all $i$.
	\item The geodesic, $\gamma$, is \emph{imprisoned} if there is a 
	compact set \cmptset{K} such that the entire image, $\gamma((a,b))$, is contained 
	in \cmptset{K}.
    \end{enumerate}
\end{defn}

Essentially the definitions differ in that imprisonment implies that 
there is some compact set which encloses the entire curve 
$\gamma((a,b))$ while partial imprisonment only requires that there be 
an infinite subsequence of points which remains in the compact set.
Hence for a partially imprisoned curve which is not totally 
imprisoned, the curve must not only continually reenter the compact 
set $K$, but must also exit it an infinite number of times (see Figure 
\ref{imprisonment.fig}).
Of course, a curve which is imprisoned is also partially imprisoned.
\begin{figure}
    \centering
    \includegraphics{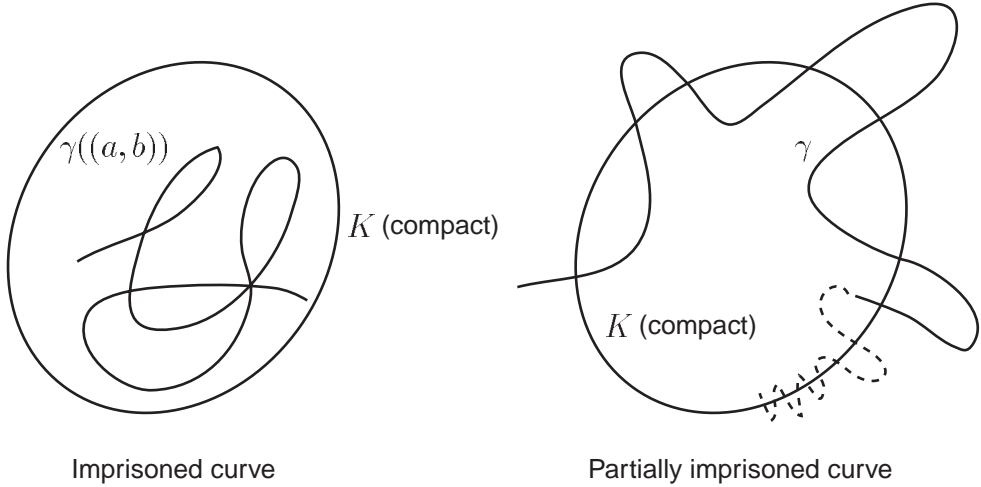}
    \caption{Imprisonment requires that the entire image of the curve 
    be contained in some compact set whereas partial imprisonment requires 
    that the curve exit and reenter a compact set an infinite number 
    of times.}
    \label{imprisonment.fig}
\end{figure}

I now present the result of Beem \cite{Beem.paper} (see also Beem \cite{BEE.book} 
p.265).
\begin{thm}
    Let \spacetime{M}{g} be a semi-Riemannian manifold.
    Assume that \spacetime{M}{g} has an endless geodesic 
    $\gamma:(a,b)\mapsto\manifold{M}$ such that $\gamma$ is 
    incomplete in the forward direction (i.e. $b\neq\infty$).
    If $\gamma$ is not partially imprisoned in any compact set as 
    $t \goesto b$, then there is a $C^{1}$-neighbourhood 
    $\nbhd{U}(g)$ of $g$ such that each $g_{1}$ in $\nbhd{U}(g)$ has 
    at least one incomplete geodesic $c$.
    Furthermore, if $\gamma$ is timelike (respectively, null, 
    spacelike) then $c$ may also be taken as timelike (respectively, 
    null, spacelike).
    \label{Beem.thm}
\end{thm}

Since strongly causal space-times do not allow past or future-directed 
non-spacelike curves to be partially imprisoned in any neighbourhood of 
a regular space-time point, Beem straightforwardly obtained the following 
corollary.

\begin{cor}
    If \spacetime{M}{g} is a strongly causal space-time which is 
    causally geodesically incomplete, then there is a 
    $C^{1}$-neighbourhood, $\nbhd{U}(g)$ of $g$, such that each $g_{1}$ in 
    $\nbhd{U}(g)$ is causally geodesically incomplete.
    \label{Beem.cor}
\end{cor}

\section{A brief introduction to $a$-boundary essential singularities}

The abstract boundary (or simply $a$-boundary) construction is a relatively 
recent addition to the collection of boundary constructions that have been 
applied to space-time.
It provides a flexible structure for the classification of boundary 
points of embeddings and appears to bypass many of the problems 
common to the $g$, $c$ and $b$-boundary constructions.
Only those definitions necessary for understanding the stability result are 
included here.
It is suggested that the reader consult Scott and Szekeres 
\cite{ScottSzekeres.paper} for a more complete and comprehensive introduction 
to the $a$-boundary construction.

In the $a$-boundary picture the boundary points in question are the 
topological boundary points, $\envbndry{\phi}{M} := \partial 
(\phi(\manifold{M}))$, of an open manifold, \manifold{M}, under the 
image of a $C^{\infty}$ embedding $\phi:\manifold{M}\mapsto 
\manifold{\widehat{M}}$.
It is important to note that both \manifold{M} and \manifold{\widehat{M}} 
are of the same dimension.
Hence $\phi(\manifold{M})$ is an open submanifold of \manifold{\widehat{M}}.
The ordered triple \envelopment{M}{\widehat{M}}{\phi} will from now on be termed 
an \emph{envelopment}.
Boundary points of different envelopments of the same manifold can turn up in 
different guises and it is useful to know when they are considered 
equivalent.
For the following we will consider two boundary sets, $\bset{B} 
\subset \envbndry{\phi}{M} \subset \manifold{\widehat{M}}$ and 
$\bset{B'} \subset \envbndry{\psi}{M}\subset\manifold{\widehat{M}'}$  from two 
different envelopments.
We say that \bset{B} \emph{covers} \bset{B'}, written $\bset{B} \covers 
\bset{B'}$ if for every neighbourhood 
$\nbhd{U}(\bset{B})\subset\manifold{\widehat{M}}$ \footnote{Note that 
$\nbhd{U}(\bset{B})$ means that \nbhd{U} is a neighbourhood of \bset{B}.} 
there exists a neighbourhood $\nbhd{U'}(\bset{B'}) \subset \manifold{\widehat{M}'}$ 
such that $\phi \circ \psi^{-1}(\nbhd{U'}\cap\psi(\manifold{M})) 
\subset \nbhd{U}$.
This definition sums up the fact that any sequence approaching \bset{B'} from 
within $\psi(\manifold{M})\subset\manifold{\widehat{M}'}$ must have 
its image sequence (i.e. mapped through $\phi \circ \psi^{-1}$) 
approach \bset{B}.
The covering relation obeys the conditions for a weak partial order 
and this leads us to the definition of equivalent boundary points, 
namely, $p \sim q$ iff $p \covers q$ and $q \covers p$.

The abstract boundary, \abndry{M}, is composed of equivalence 
classes (abstract boundary points) of boundary sets equivalent to a boundary 
point in some envelopment.
One should note that this basic structure is independent of the existence 
of a metric or chosen family of curves for the manifold and comes 
\emph{gratis}.
In order to classify abstract boundary points further we will need to 
choose a family of curves, \curvefamily{C}, obeying the \emph{bounded parameter 
property}.
The technical details of the importance of the bounded parameter 
property can be found in Scott and Szekeres \cite{ScottSzekeres.paper}.
However it suffices to say that the curves we will choose, namely, the 
family of affinely parametrised causal geodesics do satisfy this 
condition.
If there is a representative of the $a$-boundary point equivalence class 
which is the limit point\footnote{Other authors may term this an \emph{accumulation 
point} or \emph{cluster point}.} of some curve in the family, then the 
$a$-boundary point is termed \curvefamily{C}-\emph{approachable}.
This definition is internally consistent due to the formulation of 
the covering relation (see Theorem 17 of Scott and Szekeres 
\cite{ScottSzekeres.paper}).

If we provide the manifold with the additional structure of a metric 
then we can continue to classify the boundary points of an embedding by 
asking whether there exists an extension of the metric about the 
boundary point in the new envelopment.
Consequently we now assume the manifold to be endowed with a pseudo-Riemannian 
metric, $g$.
An \emph{extension} of a manifold is defined as an envelopment of a 
pseudo-Riemannian manifold, \spacetime{M}{g}, by a second pseudo-Riemannian 
manifold, \spacetime{\widehat{M}}{\hat{g}}, with embedding $\phi$ such that 
$\hat{g}|_{\phi(\manifold{M})}=(\phi^{-1})^{*}g$.
The extension will be denoted by the ordered quintuple 
\extension{M}{g}{\widehat{M}}{\hat{g}}{\phi}.
This definition simply requires that the metric over \manifold{\widehat{M}} 
agrees with the induced metric from \manifold{M} on 
$\phi(\manifold{M})$.
One should also note that this definition is consistent (but not equivalent) 
with the notion of metric extension used in Hawking and Ellis 
\cite{HE.book}.
Using this notion of metric extension we define a boundary point as 
being \emph{regular for $g$} if there exists a pseudo-Riemannian 
manifold \spacetime{\overline{M}}{\bar{g}} such that 
$\phi(\manifold{M}) \cup \{p\} \subseteq \manifold{\overline{M}} 
\subseteq \manifold{\widehat{M}}$ and \extension{M}{g}{\overline{M}} 
{\bar{g}} {\phi} is an extension of \spacetime{M}{g}.
It is important to note that unlike the notion of 
\curvefamily{C}-approachability, the regularity of an $a$-boundary point 
representative does not pass to the entire equivalence class since it 
is possible to choose poor envelopments in which representative 
boundary points are non-regular.

With these definitions in place we can now define singular boundary 
points.
\begin{defn}[singular boundary point]
    A boundary point $p\in\envbndry{\phi}{M}$ will be termed a 
    \emph{singular boundary point} or a \emph{singularity} if
    \begin{enumerate}
	\item $p$ is not a regular boundary point,
	\item $p$ is a \curvefamily{C}-approachable point, and
	\item there exists a curve, $\gamma\in\curvefamily{C}$ which 
	approaches $p$ with bounded parameter.
    \end{enumerate}
    \label{singularpoint.defn}
\end{defn}
Note that the definition of an $a$-boundary singularity is contingent 
on the choice of a suitable curve family, \curvefamily{C}.
Indeed, dependent on the choice of family, an abstract boundary point 
may be singular with respect to one family and non-singular with 
respect to another more restrictive class.
For example, if we choose \curvefamily{C} to be the family of all 
general affinely parametrised causal curves, then we would define as 
singular all those non-regular points obeying the remaining conditions 
in Definition \ref{singularpoint.defn}.
However, if we were to use the family of affinely parametrised causal 
geodesics instead, then some of those points previously defined as 
singular could now be considered non-singular.
These previously singular points, which now become non-singular, are 
those which are approachable by causal curves, but are unapproachable 
by causal geodesics.
We will term a boundary set, \bset{B}, \emph{non-singular} if none of 
its points are singular.
We are finally in a position now to define what is meant by an 
$a$-boundary essential singularity.

\begin{defn}[essential singularity]
    A singular boundary point $p$ will be termed an \emph{essential singularity} 
    if it cannot be covered by a non-singular boundary set, \bset{B}, 
    of another embedding.
\end{defn}
It is significant that despite the definition involving the concept of 
regularity, the property of being an essential singularity does pass 
through to all point members of an $a$-boundary equivalence class.
Details of this are again found in Scott and Szekeres 
\cite{ScottSzekeres.paper}.

In summary, the essential singularities we will be considering are 
(i) non-regular boundary points of an embedding, which are (ii) limit points of 
some affinely parametrised causal geodesic reached in finite 
parameter distance and which (iii) cannot be removed by the existence of a 
second embedding having non-singular boundary points covering it.
Physically this gives us the most fundamental idea of a real 
singularity as being the idealisation of a problem point of a 
space-time which is not a removable artifact and is 
`tunable'\footnote{Although it is not explicitly stated above, the 
$a$-boundary can also be tuned to the level of differentiability of 
the metric and of its extensions, as required. This will be significant later but not essential to the proof of the stability 
result.} to the incompleteness of those curves, \curvefamily{C}, considered 
physically significant.

\section{The relationship between abstract boundary singularities and 
causal geodesic incompleteness}
\label{stability.sec}

With the definitions of the former section in hand, we now present the 
result of Ashley and Scott \cite{AshleyScott.paper}.
\begin{thm}
Let \spacetime{M}{g} be a strongly causal, $C^{l}$ maximally extended, 
$C^{k}$ space-time ($1\leqslant l \leqslant k$) and $\curvefamily{C}$ be 
the family of affinely parametrised causal geodesics in 
\spacetime{M}{g}.
Then \abndry{M} contains a $C^{l}$ essential singularity iff there is 
an incomplete causal geodesic in \spacetime{M}{g}.
\label{strongcausality.thm}
\end{thm}

When considering the above theorem one must remember that it uses the 
technical definition of strong causality as defined in that paper.
This definition is consistent with that used by Beem in the 
proof of Corollary \ref{Beem.cor} and with the notions of strong 
causality presented by Hawking and Ellis \cite{HE.book} and 
Penrose \cite{Penrose.book} (see Ashley and 
Scott \cite{AshleyScott.paper}).

If we combine Corollary \ref{Beem.cor} with Theorem 
\ref{strongcausality.thm}, while taking into account the degree of 
differentiability of the metric so that the $C^{1}$-fine topology is 
well-defined, then we find the following stability result for the 
presence of $a$-boundary essential singularities.

\begin{thm}[stability of abstract boundary essential singularities]
    Suppose there exists a $C^{k}$-essential singularity in \abndry{M} for 
    a $C^{k}$ maximally extended, strongly causal space-time, 
    \spacetime{M}{g} (where $1 \leqslant k$), with family 
    \curvefamily{C} of affinely parametrised causal geodesics.
    Then there exists a $C^{1}$-fine neighbourhood, $\nbhd{U}(g)$ of $g$, so that 
    for each $g_{1}$ in $\nbhd{U}(g)$, \abndry{M} has a $C^{k}$-essential singularity 
    for \spacetime{M}{g_{1}} provided \spacetime{M}{g_{1}} is also 
    strongly causal and $C^{k}$-maximally extended for each $g_{1}$ in 
    $\nbhd{U}(g)$.
    \label{ashleystability.thm}
\end{thm}

The $C^{r}$-fine topologies defined earlier put bounds on the \emph{geometrical} 
perturbations of the metric.
Since they make no reference to the Einstein equation or stress-energy 
tensor, the metrics included in $C^{r}$-fine neighbourhoods will also 
include ones which are non-physical.
These could include, for example, geometries whose equivalent matter source 
terms violate the strong energy condition.
This is significant since energy conditions of this sort are used to 
prove the existence of incomplete timelike or null geodesics and 
consequently the existence of $a$-boundary essential singularities 
via Theorem \ref{strongcausality.thm}.
It is also possible to produce perturbations which do not coincide 
with the source of curvature for the original space-time.
For example, if a geometric perturbation is made around a vacuum 
space-time then there is no guarantee that these variations will all 
possess a vacuum source.

Minkowski space is a useful example to consider the  
$C^{r}$-stability of the inextendability of a space-time.
One perturbation that could be applied to this metric is the presence 
of small gravitational waves.
It seems unlikely for gravitational radiation of a small amplitude 
that the causal structure and maximally extended nature of Minkowski 
space would be affected.
On the other hand one could consider a Schwarzschild space-time which 
is perturbed by sending a small charge into the event horizon.
One would expect to obtain a Reissner-N\"{o}rdstrom space-time in this 
manner.
Such a case seems very physical, however, we would obtain a space-time 
that originally was maximally extended but would change its global 
structure dramatically and hence, in a physical sense, the maximally 
extended nature of the Schwarzschild space-time cannot be considered 
stable against these types of perturbation.
In the case where a neighbourhood of maximally extended metrics does 
not exist we can imagine that a perturbation of the space-time metric 
may lead to the production of sets of extension hypersurfaces for the 
space-time.

It remains an open question to show that if a space-time is strongly 
causal then that property is $C^{1}$-stable.
To the author's knowledge, this has not been proven in the literature.
Intuitively, one might expect that strong causality should be 
$C^{r}$-stable, for some $r$, since strongly causal space-times do not 
allow the existence of causal curves which leave and return to a small
neighbourhood of a manifold point.
Thus it would be expected that there exist $\epsilon$-neighbourhoods, 
whose perturbations allow causal curves passing near their own path, 
to probe out the exterior of the strong causality neighbourhoods of a 
manifold point.

\section{Concluding remarks}
At present, the abstract boundary construction proves to be the most 
promising construction with which to yield results about singularities 
in general relativity.
The above stability result guarantees the stability of the existence 
of abstract boundary essential singularities provided we also have the 
stability of the strong causality and inextendability of the 
space-time in question.
Consequently this theorem ensures the physicality of an essential 
singularity once one knows of the circumstances of its existence.
Moreover, its proof is very straightforwardly obtained.
It is hoped that the continuing stream of results involving the 
abstract boundary will bring it to the attention of Lorentzian 
geometers as a useful tool to apply to any question involving boundary 
points of space-time.

\section{Acknowledgements}
The author would like to thank Professor John Beem of the University 
of Missouri-Columbia for the initial discussion, whose questions led to this 
result.
He would also like to thank the Australian American Educational Foundation 
for the provision of a Fulbright Postgraduate Award under whose auspices 
this research was completed.
Finally he would like to thank the Mathematics and Physics Departments 
of the University of Missouri-Columbia at which the author visited while 
completing this research.


\begin{thebibliography}{99}

    \bibitem{Gerochglobalanalysis.paper}\proc{Geroch, R. and Horowitz, G. 
    T.} {Global structure of space-time} {General Relativity: An Einstein 
    Centenary Survey} {S. Hawking and W. Israel} {Cambridge: Cambridge 
    University Press} {1979} {212} {293}
    
    \bibitem{AshleyScott.paper} Ashley, M. J. S. L. and Scott, S. M., 
    \emph{in preparation}.

    \bibitem{BEE.book}\book{Beem, J. K., Ehrlich,  P. E.  and Easley, K. 
    L.} {1996} {Global Lorentzian Geometry} {Marcel Dekker Inc.} {New 
    York}

    \bibitem{HE.book}\book{Hawking, S. W.  and Ellis,  G. F. R.} {1973} 
    {The Large Scale Structure of Space-time} {Cambridge University Press} 
    {Cambridge}
    
    \bibitem{BeemEhrlich.paper}\journal{Beem, J. K. and Ehrlich, P. 
    E.} {1987} {Geodesic completeness and stability} {\MPCPS} {102} 
    {319} {328}
    
    \bibitem{Williams.thesis}\phd{Williams, P. M.} {1984} {Completeness and its 
    stability on manifolds with connection} {University of Lancaster}
    
    \bibitem{Beem.paper}\proc{Beem, J. K.} {1994} {Stability of geodesic 
    incompleteness} {Differential Geometry and Mathematical Physics, Contemporary 
    Math. Series, \textbf{170}} {J. K. Beem and K. L. Duggal} {\AMS} 
    {Providence, RI}
    
    \bibitem{ScottSzekeres.paper}\journal{Scott, S. M. and Szekeres, 
    P.}{1994}{The abstract boundary--a new approach to singularities 
    of manifolds} {\JGP} {13} {223} 
    {253}

    \bibitem{Penrose.book}\book{Penrose, R.} {1972} {Techniques of Differential 
    Topology in Relativity} {Society for Industrial and Applied Mathematics} 
    {Philadelphia}
  
\end{thebibliography}
\end{document}